\documentclass[a4paper]{jpconf}
\usepackage{graphicx}
\begin{document}
\title{Detection limit of next-generation of multi-element germanium detectors in the context of Environmental science}

\author{F.J.~Iguaz, T.~Saleem, E.~Fonda, G.~Landrot, L.~Manzanillas and F.~Orsini}

\address{Synchrotron SOLEIL, L'Orme des Merisiers, BP48 Sain-Aubin, 91192 Gif-sur-Yvette, France}

\ead{francisco-jose.iguaz-gutierrez@synchrotron-soleil.fr}

\begin{abstract}
One of the main challenges in Environmental sciences is the identification
and chemical evolution of polluting traces (e.g, cadmium or antimony) in soil,
which requires long acquistion times for accurate measurements at synchrotron facilities.
In this context, the potential of a new generation multi-element germanium detectors
to identify traces at 0.1-1~ppm in a reasonable time has been studied
using Allpix Squared framework~\cite{Spannagel:2018usc}.
This code has been customized to include
the three dimensional electric and weighting field maps generated
by COMSOL Multiphysics software, and several features to model the sample environment
at SOLEIL synchrotron and the signal response of a germanium detector
equipped with a Digital Pulse Processor (DPP).
The full simulation chain has been validated by experimental data from SAMBA beamline of SOLEIL synchrotron.
This work presents a first estimation of the detection limit to cadmium traces
in a soil sample for a future multi-element germanium detector,
using this simulation chain.
\end{abstract}

\section{Introduction}
\label{sec:intro}
The concentration of polluting trace metals in soil may be increased to toxic levels
for environmental and human health due to industrial waste, sewage and fertilizers.
However, their concentration may be still too low for an accurate chemical
identification by actual synchrotron techniques~\cite{Beccia2018}.
These techniques are limited by the sensitivity of energy dispersive detectors
to hard X-rays (energies above 20~keV), like current commercial multi-element
germanium detectors~\cite{SANGSINGKEOW2003183,Amman:2018oci},
where the maximum input count rate, the signal-to-background ratio
and the energy resolution are the key performance features.
In this context, a new generation of multi-element germanium detectors
are developed at different places to overcome actual limitations~\cite{Tartoni2020}.
In combination with the higher brillance expected for coming upgrade of synchrotron
light sources, an improvement of the detection limit is expected.

\begin{figure}[htb!]
\centering
\includegraphics[width = 0.55\textwidth]{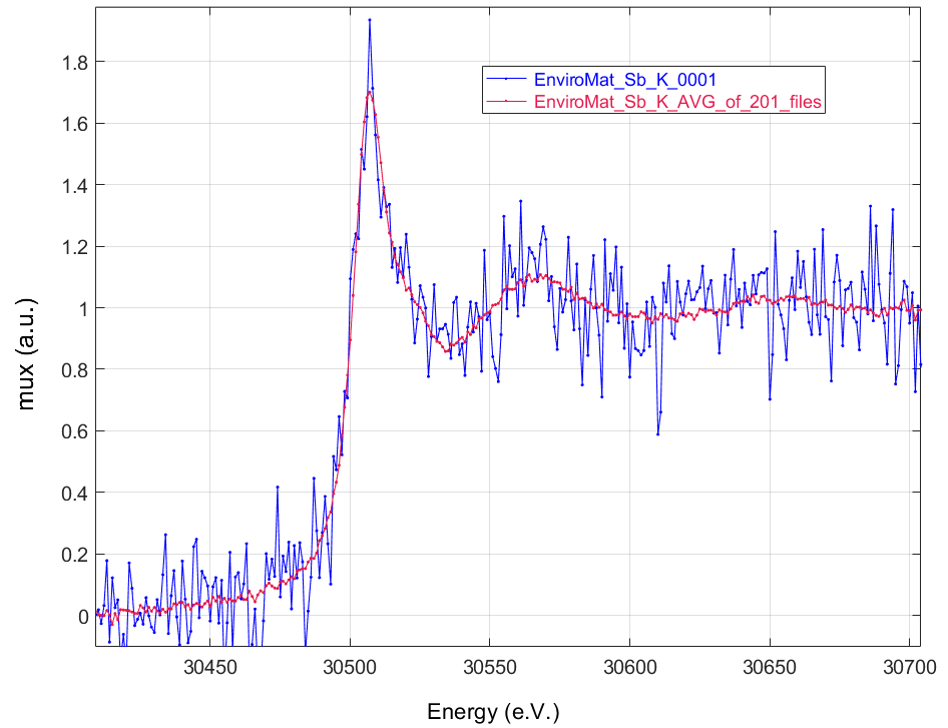}
\caption{XANES and the first part of EXAFS spectra (1 spectrum in blue, average of 201 spectra in red) of a EnviroMat standard soil.}
\label{fig:EnviroMAT}
\end{figure}

The actual sensibility of commercial multi-element germanium detectors
to polluting traces is illustrated in figure~\ref{fig:EnviroMAT},
where XANES and the first part of EXAFS spectrum of a EnviroMat standard soil
containing 5.5~ppm of antimony was measured with a 36~element germanium detector
connected to a XIA-DXP-xMAP DPP~\cite{XIAXMAP}
at the SAMBA beamline~\cite{SAMBA} of SOLEIL synchrotron.
The signal-to-noise of the 201 spectra average (red line, integration time of 5~hours)
is better than one spectrum (blue line, integration time of 95~sec),
indicating that spectra are dominated by photon counting statistics
and that an improvement of the detector throughput of a factor 10-100
without compromising the energy resolution could improve the detection limit.

In this measurement, the detector was used without any additional filtering or slitting,
perpendicular with respect to beam axis.
The sample, in the form of a pellet of compacted powder
(85~mg, 6~mm diameter, 1.25~mm thickness),
was placed at an angle of 45~degrees between the beam and the detector.
The detector-to-sample distance was of 100~mm,
corresponding to an input count rate of 42~kcps and a dead time of~20\% for all elements.
The detector could have been approached more, but the detector would have suffered of an excessive dead-time and energy resolution degradation.
The beam was focused in a small area at the sample spot (200~$\mu$m width in horizontal
and vertical axis), with a photon flux of $5.2 \times 10^{10}$~ph/sec, measured by the ionization chamber situated just before the sample.
The energy of the beam was varied between 30250 and 31200~eV
(energy steps of 1~eV, exposure time of 0.1~sec).

This work presents a first estimation of the detection limit to cadmium, one example of polluting traces in soil, for future multi-element germanium detectors based on simulation tools. Simulation tools are presented in section~\ref{sec:sim},
followed by the description of the simulation study in section~\ref{sec:meth}
and the results in section~\ref{sec:res}.
The conclusions and an outlook of future simulation studies
in section~\ref{sec:conc} complete this manuscript.

\section{Simulation of the performance of a multi-element germanium detector}
\label{sec:sim}
\begin{figure}[htb!]
\centering
\includegraphics[width = \textwidth]{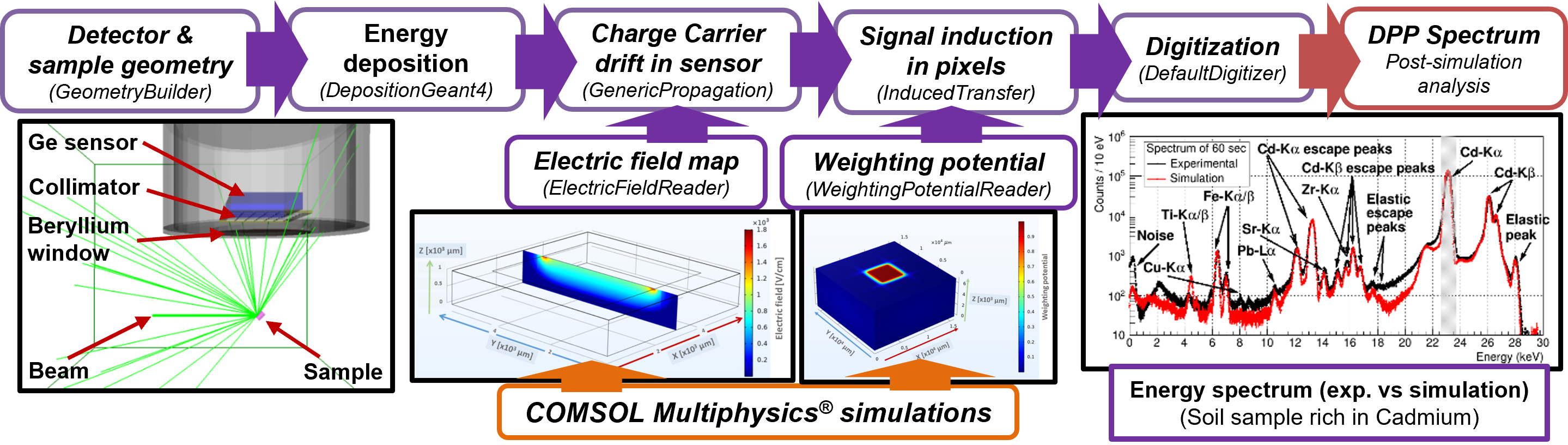}
\caption{Scheme of the full simulation chain of multi-element germanium detectors.}
\label{fig:SimChain}
\end{figure}

A first complete and fully operational simulation chain based on Allpix Squared
framework~\cite{Spannagel:2018usc} has been built,
customized to multi-element germanium detectors,
and combined with three-dimensional simulations of the electric field
and the weighting potential, based on COMSOL Multiphysics®~\cite{comsol}.
A scheme of the different simulation steps and corresponding module in Allpix Squared framework are shown in figure~\ref{fig:SimChain}.
More details can be found in~\cite{Saleem:2021mfp}.

A concise description of the different simulation steps is made for completeness:
\begin{enumerate}
 \item \textbf{Detector and sample geometry:}
 The geometry of a commercial germanium detector
 and the sample environment of SAMBA beamline is built.
 In figure~\ref{fig:SimChain} (bottom-left),
 a visualization of this geometry is shown,
 where the X-ray beam, the sample and the detector (composed of
 a beryllium window, a titanium collimator and a germanium sensor) are highlighted.
 \item \textbf{Energy deposition:}
 The interaction of X-rays with the detector geometry
 is simulated by Geant4 library~\cite{Agostinelli:2002hh}
 and energy deposits in the sensor are translated into charge carriers.
 \item \textbf{COMSOL Multiphysics® simulations:}
 Three-dimensional maps of the electric field and weighting potential
 in a germanium sensor model are simulated using COMSOL Multiphysics®.
 The sensor model includes the dimensions of all contacts and doping profiles.
 An example is shown in figure~\ref{fig:SimChain} (bottom-center).
 These maps are imported to Allpix squared by two specific modules.
 \item \textbf{Charge carrier drift in sensor:}
 Charge carriers drift along the electric field lines
 from the germanium sensor frontside to the segmented backside.
 \item \textbf{Signal induction in pixels:}
 Charge carriers movement induce signals
 in detector elements, using the Shockley-Ramo's Theorem and the weighting potential.
 \item \textbf{Digitization:}
 Induced signals on pixels are transformed into digital signals,
 using a detector noise model.
 \item \textbf{DPP spectrum:}
 Digitized signals are collected into a full energy spectrum for each pixel,
 as shown in figure~\ref{fig:SimChain} (bottom-right),
 by modeling in offline analysis a DPP~\cite{Bordessoule2019}.
\end{enumerate}

\section{Detection limit to cadmium traces in soil of a XAFS experiment}
\subsection{Methodology}
\label{sec:meth}
The detection limit of a future multi-element germanium detector
to cadmium traces in soil has been studied with simulation tools
considering a soil sample measured at SAMBA beamline~\cite{Saleem:2021mfp}.
The germanium detector modeled in this work has an active area of 900~mm$^2$
and is divided in smaller elements than actual ones
(225 elements of 4~mm$^2$ size, instead of 36 elements of 25~mm$^2$ size)
to increase the maximum input count rate.
The detector is equipped of a DPP capable to remove charge sharing events,
instead of using a collimator.
For comparison purposes, detectors with actual element size
or a recent DPP with no charge sharing rejection feature have been also studied.

\begin{figure}[htb!]
\centering
\includegraphics[width = 0.55\textwidth]{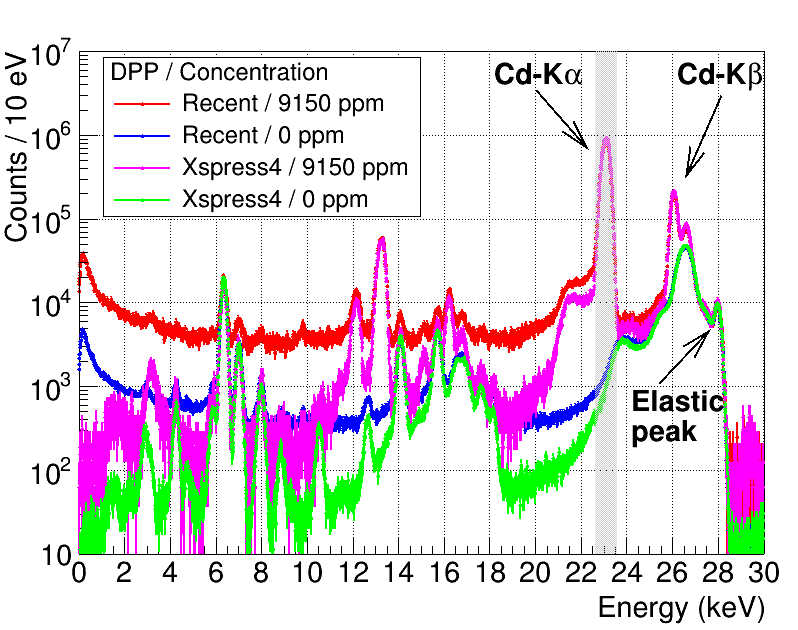}
\caption{Simulated element energy spectrum for a multi-element germanium
detector situated at 60~mm of a soil sample rich in cadmium (9150~ppm)
or of a soil with no cadmium trace.
The detector is equipped with a recent DPP or Xspress4,
and the exposure time is of 60~sec. The beam energy is 28~keV
and the beam flux is 3.47~$10^{10}$~ph/sec.
The shaded grey area defines the region of interest of the K$\alpha$-line of cadmium.}
\label{fig:EnergySpec}
\end{figure}

Charge sharing events are events where X-ray energy is shared between several elements,
involving an intensity reduction of the main fluoresence line,
an increase of the background level
and finally, a degradation of the energy resolution.
The two first effects are illustrated in figure~\ref{fig:EnergySpec}.
The main source of charge sharing events are split events,
where charge carries created by X-rays at the frontside spread over multiple elements
when drifting to detector backside.
These events are removed in current commercial detectors using a collimator
in front of the germanium sensor, at a cost of a significant reduction of the active area.
Xspress4 DPP~\cite{Xspress4} will include a firmware option
to reject charge sharing events, which is not available in recent commercial DPPs,
like XIA-FalconX~\cite{XIAFalcon}
or Xspress3M/X~\cite{QuantumXspress3,FARROW1995567}.

The detection limit to cadmium at 90\% C.L. can be expressed as
\begin{equation}
  DL (ppm) = \frac{3 \times C(ppm)}{\sqrt{N_{Pixels} \times ICR_{Pixel} \times \left(1 - \frac{DT(\%)}{100.0}\right) \times P/B \times P/T \times T_{exp}}}
  \label{eq:detlimit}
\end{equation}

where $C$ is the cadmium concentration in ppm (9150 ppm),
$N_{Pixels}$ is the number of sensor elements (36 or 225),
$ICR_{Pixel}$ is the pixel input count rate in counts per second (cps),
$DT$ is the DPP dead time in \%,
$P/B$ is the Signal-to-Background ratio,
$P/T$ is the Signal-to-Total spectrum ratio
and $T_{exp}$ is the exposure time in seconds (60 sec).
Signal (cadmium trace of 9150~ppm in soil) and background (no cadmium trace) intensities
are calculated in the region of interest (RoI)
of the K$\alpha$-line of cadmium (22.6-23.6~keV),
while the total spectrum intensity is the integral of the full energy spectrum.
In the case of signal, the background intensity is subtracted from the calculated integral.

In this first study, no three-dimensional electrostatic maps of germanium sensor
have been used in the simulation.
Instead, we have supposed a linear approximation for the electric field
and a perfect signal induction, which may underestimate
the real values for $P/B$ and $P/T$ ratios. In our simulation,
the detector-to-sample distance has been varied between
20 to 300~mm and the beam flux intensity between $10^9$ and $10^{13}$~ph/sec.

\subsection{Simulation results}
\label{sec:res}
\begin{figure}[htb!]
\centering
\includegraphics[width = 0.45\textwidth]{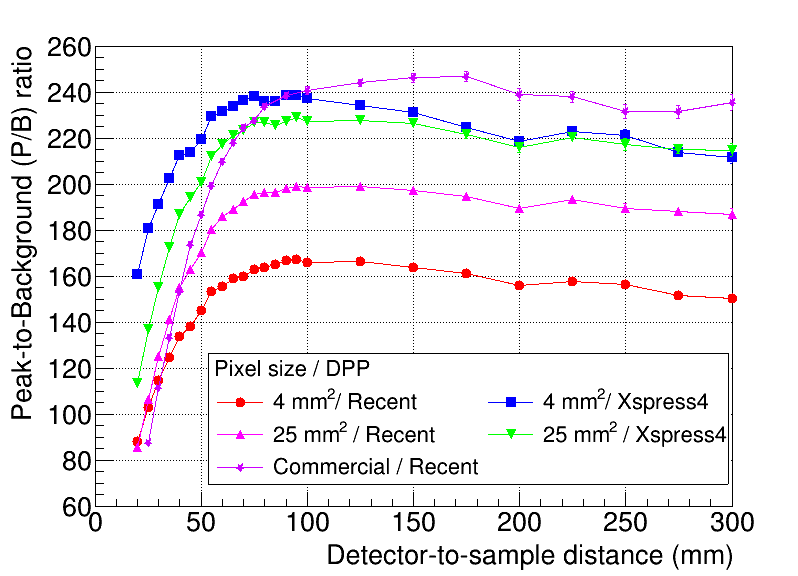}
\includegraphics[width = 0.45\textwidth]{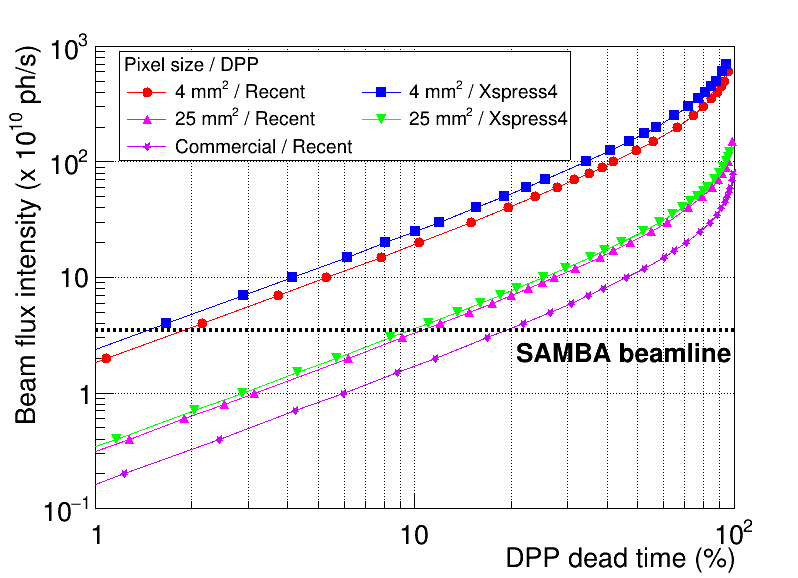}
\caption{Dependence of the peak-to-background (P/B) ratio
on the detector-to-sample distance (left)
and dependence of the beam flux intensity on the DPP dead time (right),
for a future multi-element germanium detector
with an element size of 4 or 25~mm$^2$,
and equipped with a recent DPP or Xspress4 DPP.
The estimation for an actual commercial detector
(element size of 25~mm$^2$,
equipped with a titanium collimator) is also included.}
\label{fig:PBPT}
\end{figure}

The detector-to-sample distance has been firstly optimized
studying the dependence of $P/B$ and $P/T$ ratios on the distance,
as shown for the first case in figure~\ref{fig:PBPT} (left).
$P/B$ ratio is lower for detector positions very close or very far from sample,
reaching an optimum at distance between 60 to 100~mm.
In fact, as background intensity is mainly defined by elastic beam X-rays,
this background is particularly intense if the detector is very close
to the sample. Meanwhile, if the detector is very far from the sample,
the loss of fluorescence signal is more important than
the reduction of the elastic background.
The best value for $P/B$ ratio is estimated for a detector with small pixels
and equipped with Xspress4 DPP, because the electronics will effectively remove
charge sharing events making comparable to actual commercial detector.

Then, the detector-to-sample distance has been fixed to 60~mm and
the dependence of the beam flux intensity on the DPP dead-time has been studied,
as shown in figure~\ref{fig:PBPT} (right). SAMBA beamline intensity is marked
with a dashed line. For a dead time of 20\%, a future germanium detector
could operate at beam flux a factor 2 higher than actual detector
for an element size of 25~mm$^2$, and a factor 15 for a size of 4~mm$^2$.

\begin{figure}[htb!]
\centering
\includegraphics[width = 0.6\textwidth]{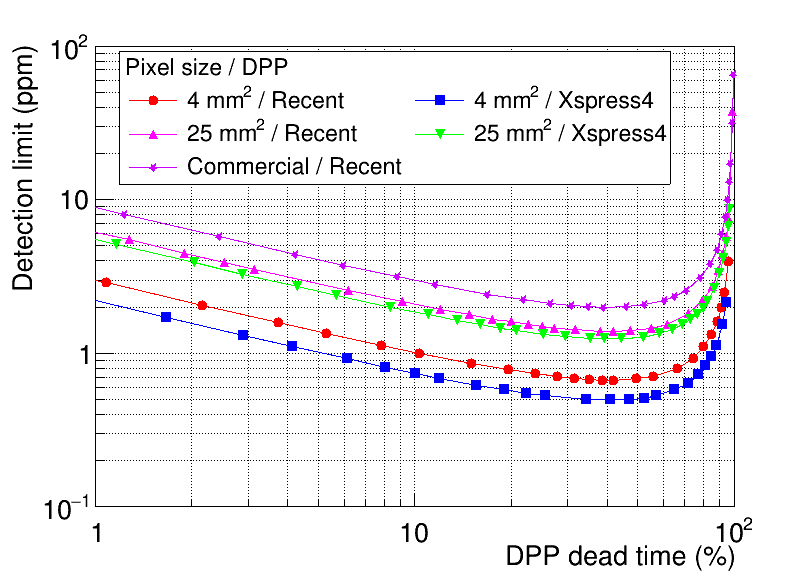}
\caption{Dependence of the detection limit to cadmium on the DPP dead time,
for a future multi-element germanium detector
with an element size of 4 or 25~mm$^2$,
and equipped with a recent DPP or Xspress4 DPP.
The limit for an actual commercial detector (element size of 25~mm$^2$,
equipped with a titanium collimator) is also included.}
\label{fig:DetLimit}
\end{figure}

Finally, the dependence of the detection limit on the DPP dead-time
is shown in figure~\ref{fig:DetLimit} for the four detector configurations.
A detection limit of 0.5~ppm to cadmium is estimated for an element size of 4 mm$^2$
and 1.25~ppm for a size of 25~mm$^2$, if Xspress4 DPP is used.
For the small element, this is a factor 4 improvement compared
to commercial detector limit (2.0~ppm), which translates in a reduction
of a factor 16 in exposure time.

\section{Conclusions and outlook}
\label{sec:conc}
A future multi-element germanium detector with very small elements
and equipped with a Xspress4 Digital Pulse Processor could increase
the sensitivity to the detection of polluting traces in soil.
This estimation is based in a customized version of Allpix Squared framework
combined with COMSOL Multiphysics software,
a complete and powerful tool that could drive current developments
of multi-element germanium detectors.
Future work to simulate the detailed three-dimensional
electrostatics maps of a germanium sensor in COMSOL Multiphysics
and SolidStateDetectors~\cite{Abt_2021} is under progress.
Alternative pixel configurations, like hexagonal ones,
are also under study to be integrated in a future release
of Allpix Squared code~\cite{Spannagel:2021tqu}.

\ack
The authors would like to thank SOLEIL computing service (ISI), in particular to Ph. Martinez, for technical support in the use of SUMO and TGCC (COBALT and TOPAZE) IT clusters.

\section*{References}
\bibliographystyle{JHEP}
\bibliography{20220404_FJIguaz_SRI2021}
\end{document}